\documentclass[prd,twocolumn,amsmath,showpacs,nofootinbib,superscriptaddress,preprintnumbers]{revtex4}
\usepackage{amsmath}
\usepackage{txfonts}
\usepackage{bm}

\usepackage{xspace}
\usepackage{multirow}
\usepackage{rotating}
\usepackage{amssymb}
\usepackage{myaasmacros}
\usepackage{comment}
\usepackage{color}

\newcommand{\dd}{\mathrm{d}}
\newcommand{\tinitial}{t_{\mathrm{initial}}}

\newcommand{\Mpc}{\mathrm{Mpc}}
\newcommand{\kpc}{\mathrm{kpc}}
\newcommand{\Msol}{\mathrm{M}_{\odot}}
\newcommand{\kNL}{k_{\mathrm{NL}}}
\newcommand{\vq}{\vec{q}} \newcommand{\vk}{\vec{k}}
\newcommand{\vx}{\vec{x}} \newcommand{\vPsi}{\vec{\Psi}}
\renewcommand{\vr}{\vec{r}}

\renewcommand{\vec}[1]{\boldsymbol{\mathbf{#1}}}

\newcommand{\hvp}[1]{}

\begin{document}

\title{Inverted initial conditions: exploring the growth of cosmic structure and voids}

\author{Andrew Pontzen}
\email{a.pontzen@ucl.ac.uk}
\affiliation{Department of Physics and Astronomy, University College London, Gower Street, London, WC1E 6BT, UK}
\author{An\v{z}e Slosar}
\affiliation{Brookhaven National Laboratory, Upton, NY 11973, USA}
\author{Nina Roth}
\affiliation{Department of Physics and Astronomy, University College London, Gower Street, London, WC1E 6BT, UK}
\author{Hiranya V. Peiris}
\affiliation{Department of Physics and Astronomy, University College London, Gower Street, London, WC1E 6BT, UK}

\date{\today}

\begin{abstract}
 
  We introduce and explore ``paired'' cosmological simulations. A pair
  consists of an A and B simulation with initial conditions related by
  the inversion
  $\delta_A(\vec{x},\tinitial) = -\delta_B(\vec{x},\tinitial)$
  (underdensities substituted for overdensities and vice versa).  We
  argue that the technique is valuable for improving our understanding
  of cosmic structure formation.  The A and B fields are by definition
  equally likely draws from $\Lambda$CDM initial conditions, and in
  the linear regime evolve identically up to the overall sign. As
  non-linear evolution takes hold, a region that collapses to form a
  halo in simulation A will tend to expand to create a void in
  simulation B.  Applications include (i) contrasting the growth of
  A-halos and B-voids to test excursion-set theories of structure
  formation; (ii) cross-correlating the density field of the A and B
  universes as a novel test for perturbation theory; and (iii)
  canceling error terms by averaging power spectra between the two
  boxes.  Generalizations of the method to more elaborate field
  transformations are suggested. 
\end{abstract}

\pacs{---}

\preprint{}
\maketitle

\section{Introduction}

The interpretation of cosmological observations increasingly requires
a precise understanding of non-linear structure formation. In addition
to the power spectrum of the matter distribution, the properties and
abundances of non-linear structures such as clusters
\cite{2015PlanckClusterSZcosmology} or voids \cite{lavaux2012voids}
also have the potential to constrain cosmological parameters.

Excursion-set theories
\cite{1974ApJ...187..425P,Bond91excursionSetHalos,Zentner07excursionSetHalos}
suggest that the formation of voids from initial underdensities is
nearly but not precisely analogous to the formation of halos from
overdensities
\cite{ShethVanDeWeygaert04voids,FurlanettoPiran06voids,Paranjape12voids,jennings2013voids}. The
imperfect symmetry suggests that directly contrasting void and halo
formation could be informative.  In this work we take a first step in
this direction by comparing results from two simulations with
precisely opposite initial conditions (underdensities substituted for
overdensities and vice versa). We refer to these simulations as being
``paired''.

The paired simulations can also be used to improve both practical
estimation and theoretical understanding of the matter power spectrum
(and higher order correlations).  There are presently two approaches
to calculating the non-linear power spectrum: analytic perturbation
theory, or computational N-body simulations. The former comes in a
wide variety of flavours, because the simplest perturbative treatment
of gravitational instability (standard perturbation theory, SPT
\cite{Bernardeau02SPTreview}) suffers from divergences at increasing
comoving wavenumber $k$.  These can be brought under control by
partially resumming some of the SPT series
\cite{CrocceScoccimarro06RPT} or writing down an effective theory
\cite{Carrasco2012EFT}. The resulting theories can be tested or
calibrated on simulations \cite{Nishimichi09BAOmodels,Copter09,
  2011MNRAS.415..829R,
  GilMartin2012SPTvssims,VlahSeljak2015LPT,Foreman15EFTLSS}.

The most familiar example of a non-standard perturbation theory is the
Zel'dovich approximation, a linear expansion in Lagrangian space which
leads to a regrouping of terms. While the raw Zel'dovich predictions
for the auto-power spectrum are inaccurate, in many respects it
behaves better than Eulerian perturbation theory
\cite{Buchert92LPT,CrocceScoccimarro06RPT,Matsubara08LPT}. In
particular, it correctly predicts the decay of the cross-correlation
between initial conditions and the final non-linear field 
\cite{CrocceScoccimarro06RPT,Bernardeau08,Matsubara08LPT,Copter09,White14Zeldovich}.
In the present work we cross-correlate the non-linear density fields
of the paired simulations, providing an alternative performance comparison of
different perturbative schemes from a physical perspective. We find that 
the Zel'dovich approximation continues to offer insight in this
new regime.

From a purely practical perspective the science case for forthcoming
large scale structure surveys requires percent-level accuracy on
computations even on strongly non-linear, megaparsec scales
\cite{Schneider15PSaccuracy}. Our third application for paired
simulations shows how they can be used to cancel a large class of
finite-volume errors that can compromise this requirement. The same
cancellation can be approximately achieved by averaging over a large
ensemble of uncorrelated simulations, but the paired approach is more
computationally efficient.

\begin{figure*}
  \includegraphics[width=1.0\textwidth]{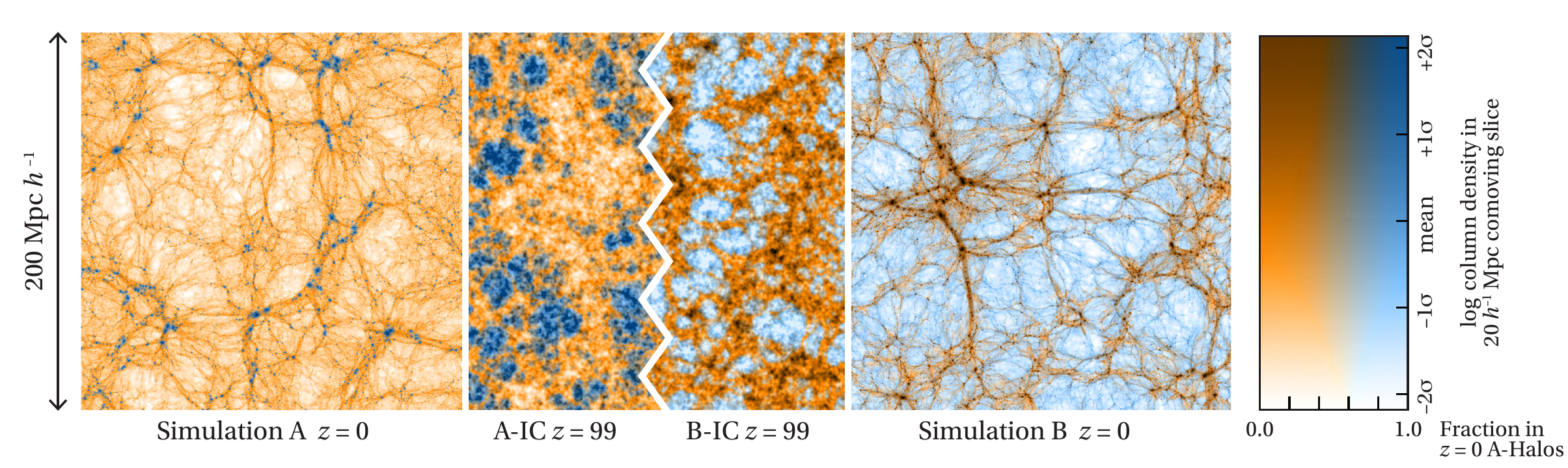}
  \caption{An illustration of the paired simulation technique. A
    standard $\Lambda$CDM simulation is performed as described in the
    text. The left panel shows the present day ($z=0$) projected
    density field in a $20\, h^{-1}\ \Mpc$ slice through the
    simulation.  Collapsed dark matter halos have been identified
    using a friends-of-friends algorithm; the fraction of the column
    density contributed by particles in such structures is
    color-coded from orange (no contribution) to blue (100\%
    contribution). The center-left panel shows the initial conditions,
    color-coded according to the same scheme. The initial conditions
    for the B simulation are obtained by reversing the sign of the
    overdensity field (center-right panel). While the statistical
    properties of the linear field are unchanged by this
    transformation, the blue ``A-halo'' particles are now associated
    with underdensities. Evolving the B-simulation to $z=0$ gives rise
    to the right panel. The B-voids are seen to be associated with the
    same particles (i.e., the same Lagrangian regions) as the
    A-halos. }\label{fig:reverse-simulation}
\end{figure*}

After describing the simulation setup (Section \ref{sec:results}) we
discuss the asymmetry in the evolution of halos and voids (Section
\ref{sec:evolution-anti-halos}) and then show how the technique
generates new insights into perturbation theory (Section
\ref{sec:pert-theory}) and improves the accuracy of power spectrum
estimates (Section \ref{sec:improving-simulations}). Possible
extensions to the technique are discussed in Section
\ref{sec:extensions}. We conclude in Section \ref{sec:conclusions}.

\section{Results}\label{sec:results}

In this paper we present results from paired cosmological simulations
drawn from an ensemble described by the WMAP 7-year recommended
cosmological parameters \cite{WMAP7cosmo} (``WMAP+BAO+$H_0$
ML''). While these are no longer the most precise parameters available
\cite{Planck15CosmoPars}, they are sufficiently close for our present
purposes where we do not compare to observational data; adopting
WMAP7 parameters allowed us to make use of an existing simulation
which we refer to as ``A''. We perform dark-matter-only simulations,
adding the baryon density to that of the dark matter.

We used CAMB \citep{Lewis:1999bs} to generate the initial power
spectrum of fluctuations from which we drew a random realization
$\vec{\delta}_S(x,\tinitial)$ on a uniform $512^3$ grid in a
$(200\,h^{-1}\,\Mpc)^3$ volume, so probing wavenumbers $0.031 < k / (h\ \Mpc^{-1}) < 16$. The initial particle displacements and velocities
were generated using the Zel'dovich approximation at redshift
$z=99$, deep in the linear regime for the relevant scales. All
simulations were run to $z=0$ using the {\sc P-Gadget-3} code
\citep{Springel:2005mi,Springel08Aquarius}. The particle softening was set to $\epsilon =
5\,h^{-1}\,\kpc$. Halos were identified with the \textsc{subfind}
algorithm \cite{Springel01SubFind}.

We generated the initial conditions for the first simulation (denoted ``A") 
using the same code as Ref. \cite{2015arXiv150407250R} and flipped the sign
of the overdensity to generate the ``B'' initial conditions, 
$\vec{\delta}_B(x,\tinitial) = -\vec{\delta}_A(x,\tinitial)$. 
Both simulations are on an equal footing in the sense that they are equally
probable draws from the underlying statistical description of the
initial conditions. However, their relationship with each other allows
for systematic investigations into various aspects of structure
formation as we describe in the following sections.

\subsection{Evolution of anti-halos}\label{sec:evolution-anti-halos}

To begin our study of the relationship between the A and B
simulations, we show that halos map reliably onto voids (and vice
versa). This situation is illustrated in
Fig.~\ref{fig:reverse-simulation} which shows, from left to right, a
$20\,h^{-1}\,\Mpc$ slice through the matter density field of the $z=0$
A-simulation, the $z=99$ A-simulation, the $z=99$ B-simulation and the
$z=0$ B-simulation. The brightness represents projected mass density while
colors track the fraction of particles identified as halos in the
A-simulation. The voids in the B simulation are identified as
A-halo-dominated regions (colored blue) interspersed by filaments,
which are A-halo-free regions (colored orange).  This relationship is
symmetric: a similar figure can be made starting from the halos in the
B simulation.

Figure \ref{fig:reverse-simulation} suggests that voids in the B
simulation can be identified with ``anti-halos'', i.e., the Lagrangian
region defined by the particles making up A-halos. However, theories
of structure formation using the excursion set formalism emphasise
that there is an asymmetry between the evolution of halos and voids
\cite{ShethVanDeWeygaert04voids,FurlanettoPiran06voids,Paranjape12voids}:
voids can be crushed by a large-scale overdensity that collapses at
late times, whereas halos are not erased by living in a large-scale
void.

The A/B comparison allows us to search for direct evidence of this
asymmetry. We take the $z=0$ A-halos in three mass bins:
$10^{14}<M/M_{\odot}<10^{15}$, $10^{13}<M/M_{\odot}<10^{14}$ and
$10^{12}<M/M_{\odot}<10^{13}$; there are respectively $353$, $4\,699$
and $38\,537$ in each range. For each A-halo at $z=0$, we track the
constituent particles through time in both A and B simulations to
follow the collapse of the halo (A) or expansion of the anti-halo (B).
At each output timestep, we record the volume-weighted\footnote{The
  volume-weighting is crucial because much of the mass within voids is
  contained inside rare but dense halos
  \cite{ShethVanDeWeygaert04voids} which contaminate the mass-weighted
  mean.} mean density of each Lagrangian region,
\begin{equation}
\langle \rho \rangle_V \equiv \frac{\sum_i \rho_i h_i^3}{\sum_i h_i^3}\textrm{,}
\end{equation}
where the sum is over all particles $i$ associated with a particular
region, $\rho_i$ is a local density estimate computed by
\textsc{pynbody} \cite{2013ascl.soft05002P} using the 64
nearest-neighbour particles, and $h_i$ is the physical distance to the
furthest of these neighbours.  We divide by the cosmic mean density
$\bar{\rho}$ to remove the effects of the background expansion. 

\begin{figure}
\includegraphics[width=0.5\textwidth]{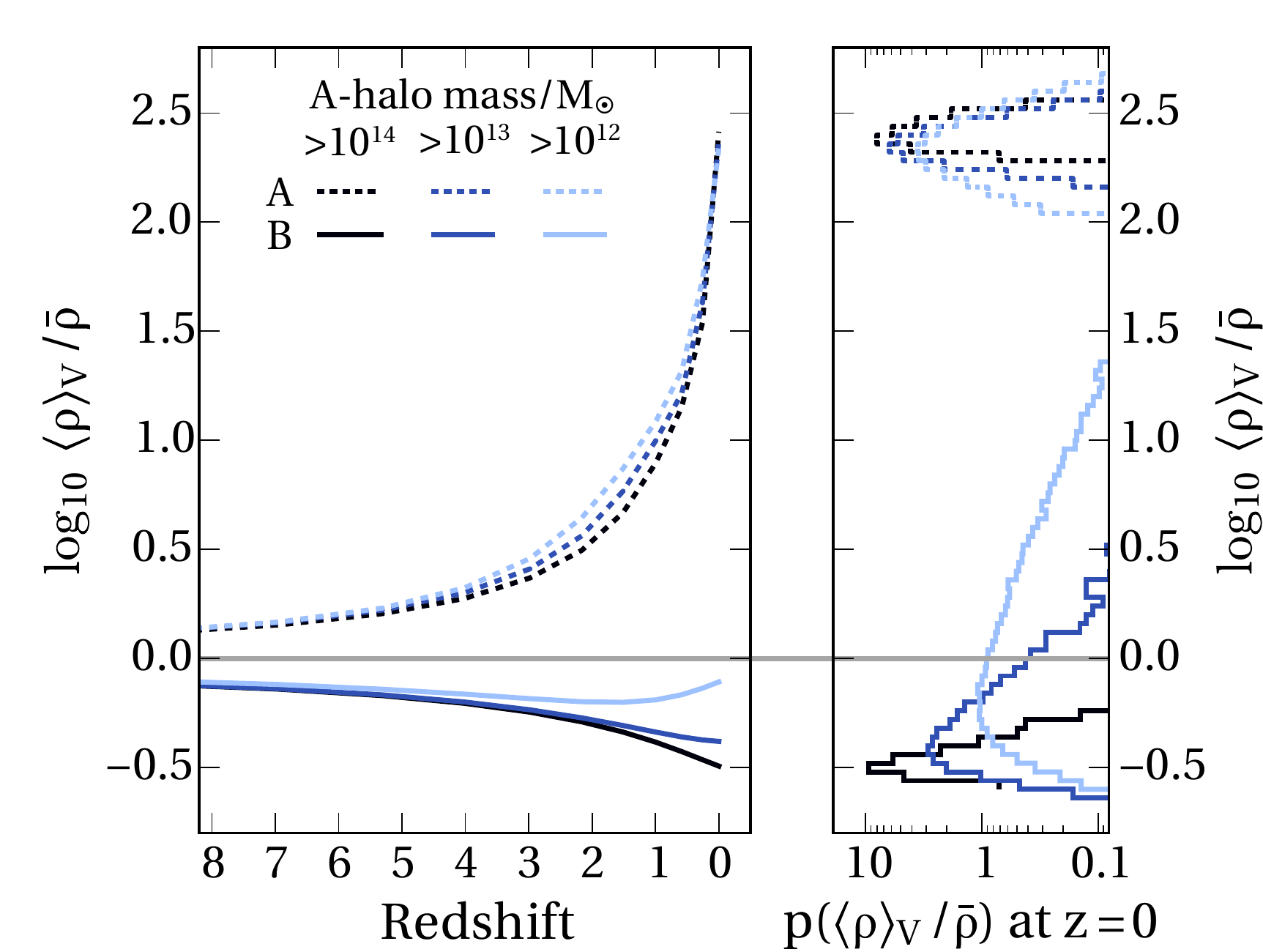}
\caption{The volume-averaged density of Lagrangian regions
  corresponding to $z=0$ A-halos of different mass ranges (from
  darkest to lightest: $10^{14}<M/M_{\odot}<10^{15}$,
  $10^{13}<M/M_{\odot}<10^{14}$ and $10^{12}<M/M_{\odot}<10^{13}$
  respectively). The left panel shows the evolution of each mass bin's
  mean density; the right panel shows the spread of halo-averaged
  densities within the bin at $z=0$. Dashed lines show the regions in
  the A-simulation (so at $z=0$ these correspond to halos); solid
  lines show the corresponding ``anti-halo'' regions in the
  B-simulation. The most massive anti-halos can be identified as
  voids. At lower masses there is a tail of crushed, high-density anti-halos,
  reflecting the known void-in-cloud evolution asymmetry. }\label{fig:antihalo}
\end{figure}

The results of the density calculation are shown in Fig.~\ref{fig:antihalo}. Over time, the Lagrangian
region corresponding to the final $z=0$ halos grows in density (dashed lines, left panel). The
differences between the three mass bins are relatively small, with a
slight trend for lower-mass regions to reach higher densities at earlier
times. The right panel shows a histogram of the densities of the
individual halos making up each mass bin at $z=0$; once again, the
A-simulation results are shown by dashed lines. The
variance in the mean density is small, which is to be expected given
that the halos are identified based on a friends-of-friends
algorithm which specifies a fixed density for their boundary
\cite{Springel01SubFind}.

The solid lines show the corresponding quantities for the anti-halo
regions in the B-simulation. The left panel shows that, at early
times, the selected regions are underdense, as demanded by the
antisymmetry in the initial conditions. Over time the largest
anti-halos become progressively less dense, as expected for voids. The
histogram (right panel) confirms that the most massive 
anti-halos are all well below the cosmic mean density and can be
robustly identified as voids, confirming the more qualitative picture
painted by Fig.~\ref{fig:reverse-simulation}.

In the lowest mass bin, the average density of the anti-halos turns
around and starts to grow (relative to the cosmic mean) at low
redshift. This is consistent with the expected ``void-crushing''
process \cite{ShethVanDeWeygaert04voids}. The right panel shows that
the majority of anti-halos remain underdense, but the mean is dragged
up by a few regions. Inspection of these high-density cases confirms
that they are being crushed by larger-scale collapse.  The effect is
only evident at low mass; otherwise, even if anti-halos are contained
within a B-overdensity, there has not been time for gravitational
collapse to crush them.  We can describe the anti-halos above a
density threshold of $\langle \rho \rangle_V/\bar{\rho}>200$ as
``fully crushed", since they have achieved a mean density comparable
to that of a halo.  Even at $10^{12}\,\Msol$ (the minimum mass we can
reliably resolve), only $0.1\%$ of anti-halos at $z=0$ exceed the
threshold. It is far more common to find anti-halos that have been
crushed only along two dimensions, and now form the diffuse mass in a
cosmic filament.

In summary, anti-halos correspond closely to voids, especially on
large scales. There is presently significant interest in formulating
reliable ways of defining voids so that such structures can be
identified and used for cosmological inference in large scale
structure surveys \cite{Sutter14VoidID}. By selecting anti-halos that
have not been crushed, one could arrive at a clean definition of
voids. We will explore this further in a future paper.

\subsection{Perturbation theory}\label{sec:pert-theory}
In this Section we show how our paired simulation approach can be used
to study cosmological perturbation theory. Given a density field for a
given simulation labeled $X$ we define
\begin{equation}
\delta_X(\vk)  \equiv \int \dd^3 \vec{x}  \,
  \left(\frac{\rho_X(\vec{x})}{\bar{\rho}} - 1 \right)\, e^{-i\vk\cdot
                \vx}\textrm{,}
\end{equation}
where $\vec{k}$ is a comoving wavevector, $\delta_{X}(\vec{k})$ is the
Fourier-space overdensity, $\vx$ is the comoving position,
$\rho_X(\vx)$ is the density, and $\bar{\rho}$ is the mean
density. The cross-power spectrum between fields $X$ and $Y$,
$P_{XY}(k)$, is defined by
\begin{equation}
  \frac{1}{2}\left\langle \delta_X^{\star}(\vec{k}) \delta_Y(\vec{k'})
    + \delta_Y^{\star}(\vec{k}) \delta_X(\vec{k'})\right\rangle  =
  (2\pi)^3 P_{XY}(k)\delta^D(\vec{k} - \vec{k}')\textrm{,}\label{eq:define-powerspec}
\end{equation}
where angular brackets denote the ensemble average, and $\delta^D$
represents the Dirac delta function.  We will make use of the simulated
density fields A and B, but also the linearly-evolved field L which is defined
as
\begin{equation}
\delta_L(\vec{k}, t) = \delta_A(\vec{k}, \tinitial) D(t,\tinitial)\textrm{,}\label{eq:linear-field-definition}
\end{equation}
where $D(t,\tinitial)$ is the linear growth factor.  From these three
fields, there are six power spectra that can be constructed; however,
in the true ensemble average, two of these ($P_{AA}=P_{BB}$ and
$P_{AL}=-P_{BL}$) contain identical information. (In practice, since
the volume of simulations is finite, there is residual information in
$P_{AA}-P_{BB}$ and $P_{AL}+P_{BL}$ that we will discuss in Section
\ref{sec:improving-simulations}.) 

We modified the \textsc{GenPK}
code\footnote{\url{http://github.com/sbird/GenPK}}
\cite{Bird11_Lya,2012MNRAS.420.2551B} to calculate cross-correlations
between two \textsc{Gadget} outputs. For the purposes of the present
discussion we construct four power spectrum estimates:
$P(k)=(P_{AA}(k)+P_{BB}(k))/2$,
$P_{\times L}(k) = (P_{AL}(k)-P_{BL}(k))/2$, $P_{AB}$ and
$P_{LL}(k)$. We assume that particle shot noise is uncorrelated
between A and B simulations and therefore do not subtract its
contribution to the cross-spectra; the validity of this assumption
does not affect our results, since shot noise is highly subdominant
over the scales of interest.

\begin{figure}
  \includegraphics[width=0.5\textwidth]{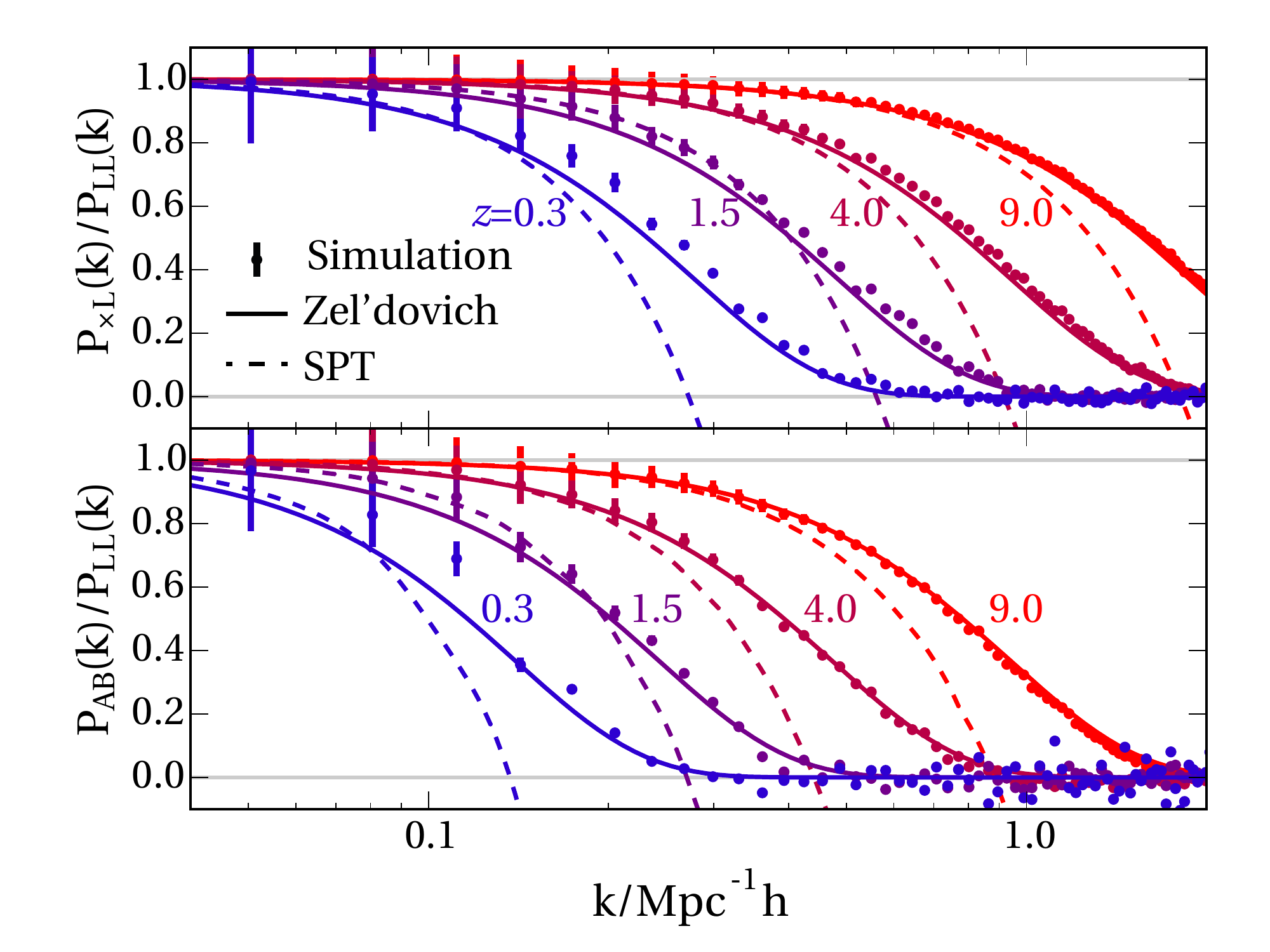}
  \caption{The cross-power spectra, for four redshifts $0.3<z<9.0$, of
    simulated and linearly-evolved fields ($P_{\times L}$, upper
    panel) and A and B simulations ($P_{AB}$, lower panel), each
    normalized by the linear power $P_{LL}$.  The Zel'dovich
    resummation from Appendix \ref{sec:zeldovich-cross-corr} is shown
    as a solid line, and gives excellent agreement with the
    simulations even at low redshift. Standard perturbation theory at
    1-loop order is shown by the dashed line and is in poor agreement
    with the simulations at all redshifts. These results are known for
    the linear cross-correlation (upper panel) but continue to hold
    for the new $P_{AB}$ cross-correlation (lower
    panel). }\label{fig:cross-ps}
\end{figure}

We start by focussing on the cross-correlations $P_{\times L}$ and
$P_{AB}$; these are plotted for a range of redshifts in
Fig.~\ref{fig:cross-ps} (upper and lower panels respectively),
normalized by $P_{LL}$. The quantity $P_{\times L}/P_{LL}$ is
sometimes called the propagator; it expresses the degree of coherence
between the non-linear and linear fields and has been studied
extensively
\cite{CrocceScoccimarro06RPT,Crocce06RPTPropagator,Matsubara08LPT,Copter09}.
These studies have revealed that the connection between initial
overdensity and final non-linear structure is poorly described by
standard perturbation theory (SPT), but (as we re-derive in Appendix
\ref{sec:zeldovich-cross-corr}) accurately predicted by resumming the
Zel'dovich approximation which gives
\begin{equation}
P_{\times L}^{\mathrm{zel}}(k) = e^{-(k/\kNL)^2}P_{LL}(k)
\end{equation}
where $\kNL$ is the wavevector corresponding to the scale at which the
linear and non-linear fields decohere,
\begin{align}
\kNL^{-2} &= \frac{1}{12\pi^2} \int_0^{\infty} P_L(k') \dd k'\textrm{.}
\end{align}
The qualitative reason for this decoherence is straight-forward: in
the non-linear evolution, the particles have moved (from their initial
positions) an r.m.s. distance $\langle \Delta x^2 \rangle^{1/2}$ that
is proportional to $\kNL^{-1}$; for scales below this limit, the
original information has been erased by the displacements. However it is unclear
why the Zel'dovich description provides such a good
quantitative fit to the propagator.\footnote{Even better agreement could be found by fitting the
    scale of the Gaussian suppression, $k_{\rm{NL}}$, at each redshift.} By contrast, finite-order standard perturbation
theory  --- an expansion in the Eulerian density contrast, shown by dashed lines in Fig.~\ref{fig:cross-ps} --- performs worse in describing the
decorrelation. For that reason, the Zel'dovich result has been used as
an inspiration for partially resumming perturbation theory to combine
the best of both worlds. For example, both resummed standard
perturbation theory (RPT)
\cite{CrocceScoccimarro06RPT,Crocce06RPTPropagator} and resummed
Lagrangian perturbation theory (LPT) \cite{Matsubara08LPT} are designed to reproduce
the Zel'dovich cross-correlation result at tree level \cite{Copter09}. 

\begin{figure*}
  \vspace{1cm}
  \includegraphics[width=0.8\textwidth]{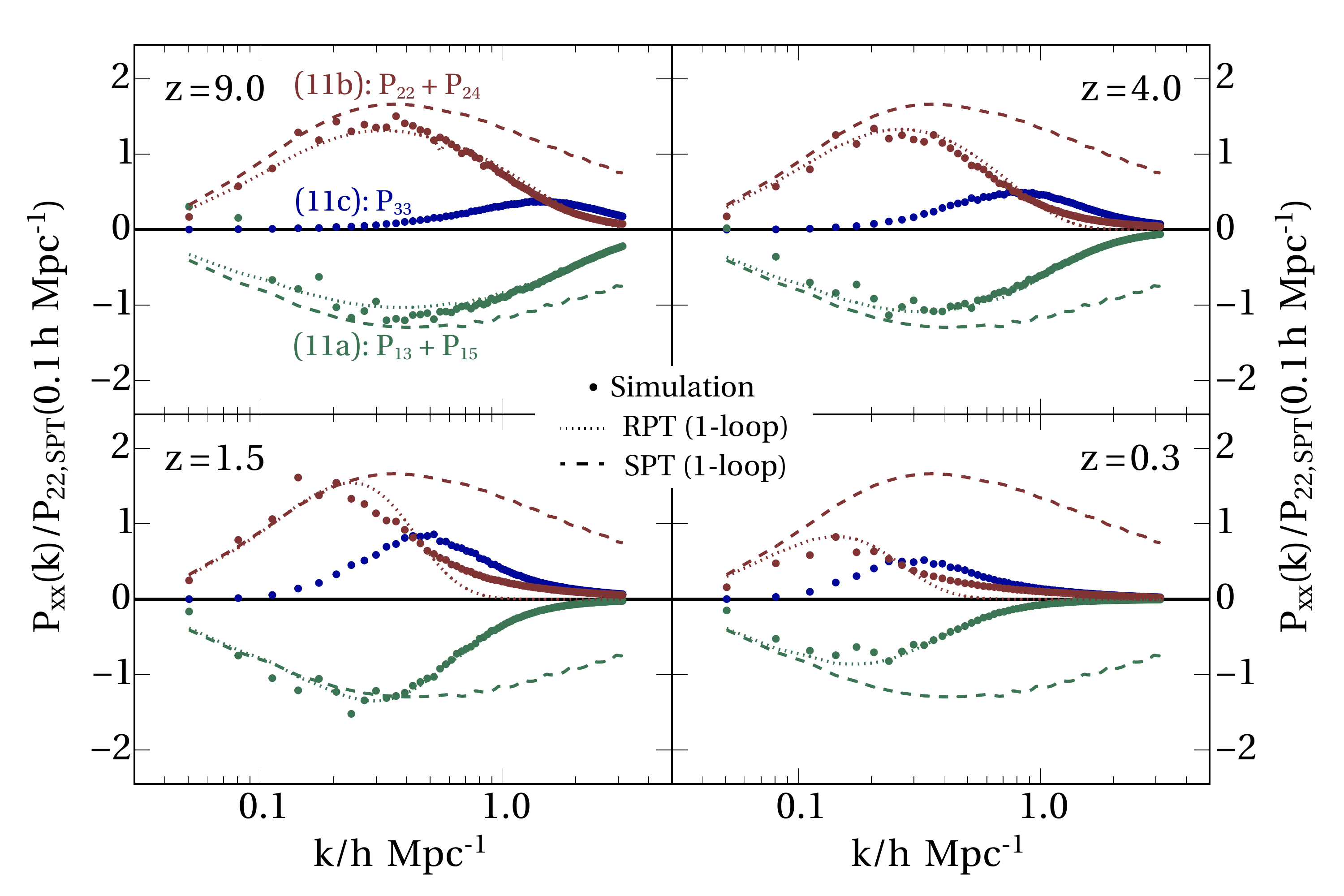}
  \caption{The power spectrum
    empirically split into different SPT series according to
    Eq.~\eqref{eq:ps-parts-inversion}, normalized for convenience to
    $P_{22,\mathrm{SPT}}(0.1\,h\,\Mpc^{-1})$.  The three equations are
    labeled by their SPT contributions.  Dashed lines show the SPT
    calculation for these terms (although we can only compute for
    1-loop order).  Dotted lines show the equivalent results for RPT
    which incorporates high-order effects even in the 1-loop
    truncation. Points show measurements from the paired simulations,
    from which we can verify the effectiveness of resummation, and
    also see directly that the magnitude of 2-loop $P_{33}$ terms is
    small at sufficiently low $k$. }\label{fig:perturbation-theory}
\end{figure*}

We will now show that the cross-correlation between the A and B simulations provides a new
testbed for perturbation theory schemes. The $P_{AB}$ measurements from the simulations are shown in
the lower panel of Fig.~\ref{fig:cross-ps} along with 1-loop SPT
(dashed lines), showing that the perturbation theory gives a
reasonable prediction at sufficiently low $k$ but once more diverges
in the high-$k$ limit\footnote{We used the \textsc{Copter} code \cite{Copter09} to calculate
  perturbation theory results for this paper.}.  As in the AL case, it is possible to use the
Zel'dovich approximation to find a far better description of the AB
decorrelation. The solid line in the lower panel of Fig.~\ref{fig:cross-ps} shows the result, derived in 
Appendix \ref{sec:zeldovich-cross-corr}:
\begin{equation}
P^{\mathrm{zel}}_{AB}(k) = e^{-(2k/\kNL)^2}P_{LL}(k)\textrm{,}\label{eq:AB-zeldovich}
\end{equation}
giving an excellent fit to the simulation results. One justification
for this form is to imagine that the particle displacements are
doubled in magnitude relative to the AL displacements, so that the
relevant wavenumber is halved. However the more formal derivation of
Eq.~\eqref{eq:AB-zeldovich} as given in Appendix
\ref{sec:zeldovich-cross-corr} requires an unconventional choice of
resummation. The reason why this particular choice (or even the
underlying Zel'dovich approximation itself) is so successful is unclear and
discussed further in the Appendix.

Instead of directly studying cross-correlations, we can use the new
information to empirically constrain the terms within SPT. In this
approach, the non-linear overdensity field $\delta_A$ is written as
the sum of terms of increasing powers of the linear field,
$\delta_A = \delta_1 + \delta_2 + \cdots$. The power spectrum is then
expanded as a series in the auto- and cross-power of these individual
terms; schematically
\begin{equation}
P_{AA} = P_{BB} =  P_{11} + P_{13} + P_{22} + P_{15} + P_{24} + P_{33} + \cdots \textrm{;} \label{eq:spt-ps}\\
\end{equation}
to two-loop order, where $P_{ij}$ denotes the parts of the power
spectrum formed from contracting a term which is $i$th order in
$\delta_L$ with one which is $j$th order.\footnote{By convention the $P_{ij}$
term absorbs the $P_{ji}$ term if $i \ne j$; this leads to a
potentially confusing factor 2 notational discrepancy between the
formal definition of symmetric ($i=j$) and asymmetric ($i \ne j$)
terms \cite{1992MakinoSPT}. We nevertheless adopt this convention for compatibility with the existing
literature. } Note that there are no terms for which $i+j$ is odd,
since the ensemble expectation value is identically zero in such
cases. We have suppressed the $k$ parameter for brevity.

By cross-correlating the A and B simulated densities with the linear
field, we obtain the series
\begin{equation}
  P_{AL} = -P_{BL} = P_{11} + \frac{1}{2} \left(P_{13} + P_{15} +
    \cdots \right)\textrm{,}\label{eq:spt-lin-ps}
\end{equation}
where the factor $1/2$ arises because, unlike in the autocorrelation
case, there are no $P_{ji}$ terms to absorb into the $P_{ij}$
terms. Cross-correlating with the B field gives
\begin{equation}
P_{AB} = -P_{11} - P_{13} + P_{22} - P_{15} + P_{24} - P_{33} + \cdots \textrm{,}\label{eq:spt-X-ps} \\
\end{equation}
where we have picked up a minus sign in front of each $P_{ij}$ term
for which $i$ and $j$ are odd (so that an odd number of B linear
fields appears).

Truncating at the two-loop order, the relationships \eqref{eq:spt-ps},
\eqref{eq:spt-lin-ps} and \eqref{eq:spt-X-ps} can be partially
inverted to obtain 
\begin{subequations}\label{eq:ps-parts-inversion}
\begin{align}
P_{13} + P_{15} &= \frac{1}{2}\left(P_{LA}-P_{LB}\right)
-P_{LL}\textrm{;}  \\
P_{22} + P_{24} &= \frac{1}{2}\left(P_{AA}+P_{BB} + P_{AB}\right)\textrm{;} \\
P_{33} &= P_{LL}-P_{LA}+P_{LB}-\frac{1}{2} P_{AB}+\frac{1}{4} \left(P_{AA} + P_{BB}\right)\textrm{.}
\end{align}
\end{subequations}
The leading-order LHS of Eqs. (\ref{eq:ps-parts-inversion}a, b) can be
computed using 1-loop SPT. The RHS can be measured from our paired
simulations, and predicted by RPT (or other resummed theories). The
comparison is shown in Fig.~\ref{fig:perturbation-theory}: the
simulation results are shown as points, SPT by dashed lines and RPT by
dotted lines.

The 1-loop SPT predictions for the $P_{13}$ and $P_{22}$ class terms
are poor for $k>0.2\,h\,\Mpc^{-1}$, 
although the errors are opposite in sign and so
significantly cancel to produce reasonable predictions of the
autocorrelation 
\cite{2006JeongKomatsuSPT}. The strength of RPT (dotted lines) in
predicting the $P_{13}$ series derives directly from its exact
agreement with the Zel'dovich approximation in the $P_{AL}$
cross-correlation (Fig.~\ref{fig:cross-ps}).

By adding the AB cross-correlation we have been able to extract
 $P_{33}$ (a two-loop term) directly from simulations for the first time. 
 Such high-order terms must be small at low $k$ for perturbation theory
 to be valid.   Our
results demonstrate that this requirement does in fact hold in numerical simulations.

Being able to extract different perturbation theory terms empirically also gives
the opportunity for testing resummation schemes. The addition of the ``B" simulation
gives access to a distinctive higher-order test that is not available from existing methods.
At present we do not have code to calculate 2-loop predictions so this
comparison is left for future work.

\subsection{Improving the accuracy of structure formation simulations}\label{sec:improving-simulations}

\begin{figure}
  \includegraphics[width=0.5\textwidth]{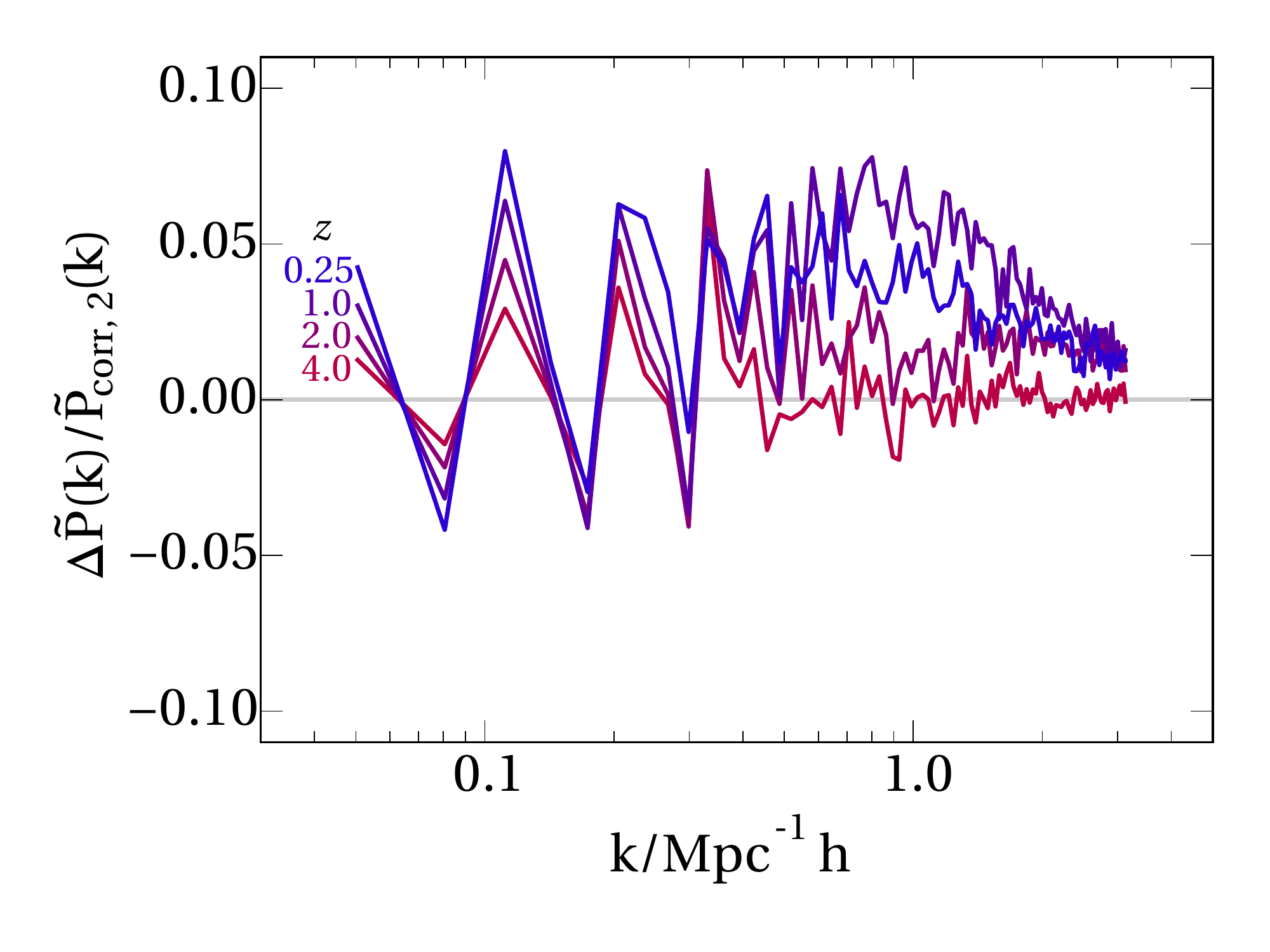}
  \caption{Finite-box errors in power spectrum estimates can be vastly
    reduced using a paired simulation. Here we show our correction for
    the next-to-leading-order error term
    $\Delta \tilde{P}(k) = \tilde{P}(k)_{\mathrm{corr,2}} -
    \tilde{P}(k)_{\mathrm{corr,1}}$
    as a fraction of $\tilde{P}(k)_{\mathrm{corr,2}}$.  Despite having
    a reasonable $200\,h^{-1}\,\Mpc$ box size, correlated artefacts
    from the small number of large-scale modes propagate down to
    create an apparent bias in the power spectrum, reaching
    $\simeq \,5\%$ on scales as small as $k=1\,\Mpc^{-1} h$ at
    $z=1$. This cannot be corrected by older techniques that divide
    out cosmic variance in the linear power, as it is an inherently
    non-linear effect. }\label{fig:volume-error}
\end{figure}

As a final example application of paired simulations, we turn to a
more immediately practical question. Since numerical simulations of
non-linear structure formation can probe only a finite dynamic range,
practitioners need to balance the box size against the ability to
resolve small scales. The finite volume has two effects: first, it
removes all power below $k_{\mathrm{min}} = 2 \pi / L_{\mathrm{box}}$,
where $L_{\mathrm{box}}$ is the comoving size of the box
\cite{Li2014SuperSample}. This could be tackled in a
computationally-efficient way by assuming a separate-universe
approximation and rescaling the background cosmology in each ``patch''
\cite{Angulo2010RescaleNbody}; we will not consider this further here.
The second effect of the small box is that it samples only a small
number of modes for wavenumbers reaching $k=k_{\mathrm{min}}$, leading
to variance effects that vanish in a true ensemble mean. Paired
simulations can be helpful in tackling this problem.

In this section we will need to make a clear distinction between the
theoretical power spectrum $P(k)$, as defined by Eq.~\eqref{eq:define-powerspec}, 
and the measured power spectrum
$\tilde{P}(k)$ which is defined with reference to the discretized
density field components in a simulation
\begin{equation}
\tilde{P}(k) = \frac{1}{N_k} \sum_{i \in S_k} \delta_i^{*} \delta_i \, ,\label{eq:power-spec-estimate}
\end{equation}
where $\delta_i$ is the density field Fourier component with index
$i$, $S_k$ is the set of such components that are used in the power
spectrum estimate for wavenumber $k$, and $N_k$ is the size of that
set. Shot-noise corrections \cite{2005ApJ...620..559J} can be applied
to Eq.~\eqref{eq:power-spec-estimate} without changing the
discussion; we omit it for simplicity. Expanding
$\delta_i$ to second order in perturbation theory, we have
\begin{equation}
\tilde{P}(k) \simeq \frac{1}{N_k} \sum_{i \in S_k} \left( \delta_{i,L}^{*} \delta_{i,L} +
G_{ijk} \left(\delta_{j,L}^{*} \delta_{k,L}^{*} \delta_{i,L}\right) + \mathrm{c.c.} +
\cdots \right)\label{eq:discrete-ps}
\end{equation}
where $\delta_{i,L}$ is the linear amplitude for component $i$,
c.c. indicates the complex conjugate of the preceding term, $D$ is the
linear growth factor, and $G_{ijk}$ describes how mode $i$ grows in
response to the amplitude of modes $j$ and $k$, evaluated at a given
redshift. The linear growth function has been absorbed into the
definition of the linear field according to Eq.~\eqref{eq:linear-field-definition}. 
There is an assumed summation over
all modes $j$ and $k$.

Given simulations of a specified box size, the ideal quantity to
calculate is $\langle \tilde{P}(k) \rangle$, i.e., an average over all
possible realizations of the initial field $\delta_{i,L}$. This is
typically attempted by computing tens or even hundreds of realizations
\cite{Heitmann10Coyote,Heitmann14CoyoteEmulation}, which is
computationally costly. The leading-order correction that this
generates compared to a single realization can actually be applied by
hand, since $\langle \tilde{P}(k)\rangle = P_L(k) + \cdots$. The
required ``first-order corrected'' power spectrum estimate is
\begin{equation}
\tilde{P}(k)_{\mathrm{corr,1}} = \tilde{P}_{AA}(k) +
P_L(k) - \tilde{P}_{LL}(k)\,\textrm{,}
\end{equation}
so that $\tilde{P}(k)_{\mathrm{corr,1}} - \langle \tilde{P}(k)
\rangle$ is third order in $\delta$.
Note that this is different from the usual approach of ``canceling''
sample variance \cite{2005Natur.435..629S} by estimating
\begin{equation}
\tilde{P}(k)_{\mathrm{corr,std}} = \tilde{P}_{AA}(k)\frac{P_{L}(k)}{\tilde{P}_{LL}(k)},
\end{equation}
which is hard to justify from a theoretical point of view (though one
gets the right answer on linear scales by construction).

With a paired simulation in hand, we can
go further and apply the next-to-leading-order correction because, inspecting
Eq.~\eqref{eq:discrete-ps}, the error has odd parity in
$\delta_L$ and so reverses sign in $P_{BB}$. Thus,
\begin{equation}
\tilde{P}(k)_{\mathrm{corr,2}} = \frac{1}{2} \left(
  \tilde{P}_{AA}(k) + \tilde{P}_{BB}(k) \right) +
P_L(k) - \tilde{P}_{LL}(k)\textrm{.}
\end{equation}
Residual errors in $\tilde{P}(k)_{\mathrm{corr,2}}$ compared to
$\langle \tilde{P}(k) \rangle$ are then fourth order in $\delta$.

The corrections arising from this change are highly significant in the
case of a $200\,h^{-1}\,\Mpc$ box. Figure \ref{fig:volume-error} shows
the correction
$\Delta \tilde{P}(k) = (\tilde{P}_{BB}(k) - \tilde{P}_{AA}(k))/2$, as
a fraction of $\tilde{P}(k)_{\mathrm{corr,2}}$. The corrections reach
$\sim \, 5\%$ even at small scales, $k=1.0 \ h^{-1} \Mpc$, and modest
redshifts, $z=1$, where one might hope the box size effects to be
minimal.  This is consistent with what is found by averaging over
hundreds of realizations \citep{Heitmann14CoyoteEmulation} or
comparing to increased box sizes which better sample the large-scale
modes \citep{Schneider15PSaccuracy}.

Moreover, unlike the second-order cosmic variance, the third-order
error is strongly correlated over different $k$'s, presumably because
it arises from the coupling to a small number of low-$k$
modes \cite{1999MeiksinWhiteCorrelatedPk}. In other words, the sample
variance is not determined solely by the number of modes in the
initial conditions at the same wavenumber, but can instead be dominated by
the $G_{ijk}$ coupling to the poorly-sampled low-$k$ modes.

By performing just one additional simulation, it is possible to
remove this bias to third order accuracy. The fourth-order term is
left invariant by the averaging. Overall, our paired technique enables
significant gains in computational efficiency when generating
non-linear power spectra for comparison with large cosmological
surveys.

\section{Extensions}\label{sec:extensions}

\begin{figure}
  \includegraphics[width=0.5\textwidth]{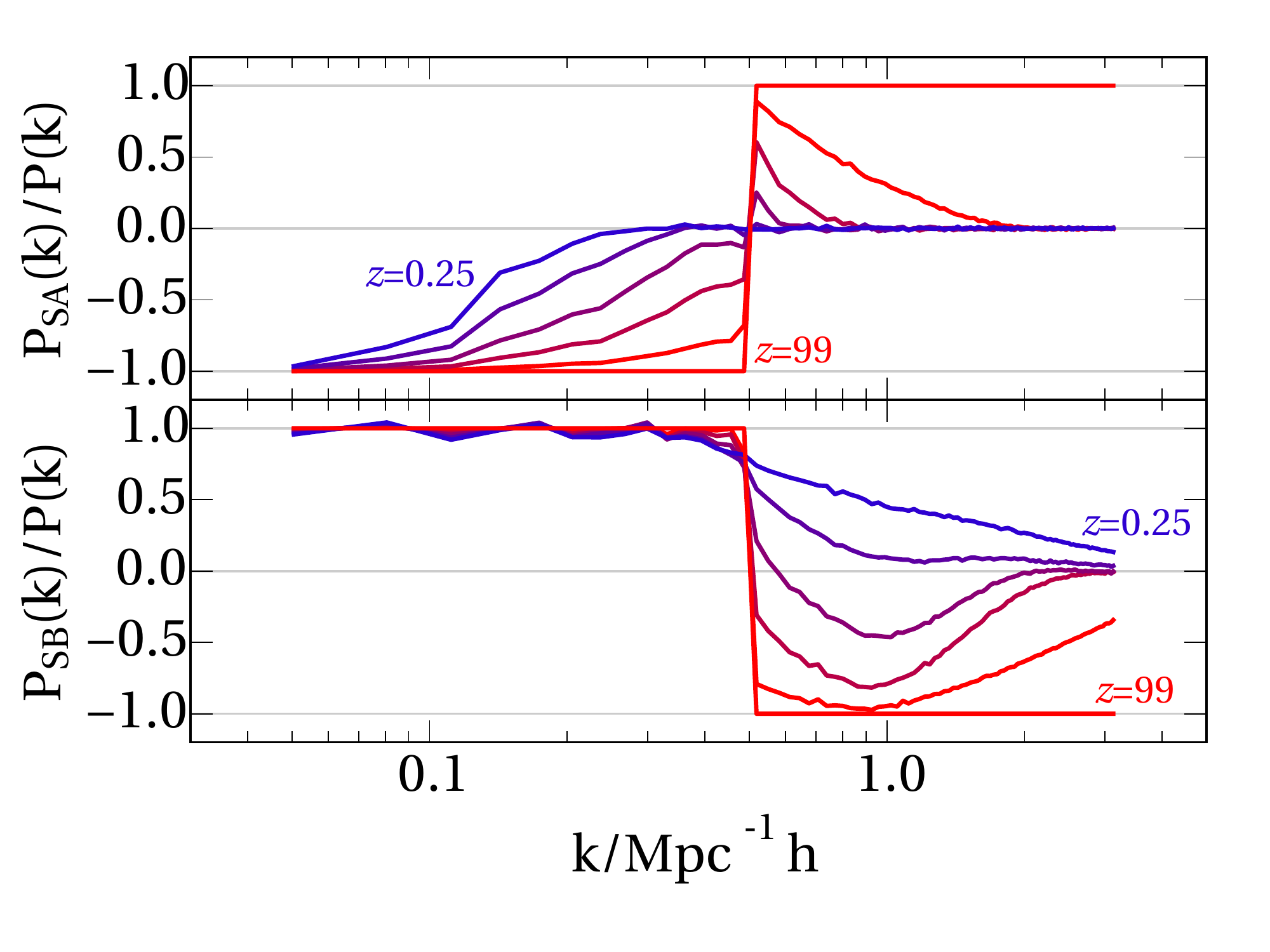}
  \caption{As an example extension, we show the
    cross-correlation between the spliced (``S'') simulation and the
    original (``A'' and ``B'') pair. Modes in the S initial conditions
    are equal to the ``A'' modes on small scales
    ($k>0.5\,\Mpc^{-1}\,h$), but to the ``B'' modes on large scales
    ($k<0.5\,\Mpc^{-1}\,h$). In cross-correlation with the A
    simulation, one sees the effect of large-scale streaming
    decorrelating the small scales. Conversely, the lower panel shows
    a trend towards coherence between S and B simulations on
    small scales, showing that the final power at large $k$ is
    being sourced by low-$k$ density fluctuations. All three
    simulations have a $\Lambda$CDM power spectrum, so the effects
    are being directly measured within a realistic cosmological
    setting rather than a toy model.}\label{fig:spliced}
\end{figure}

As well as fleshing out the three applications in Section
\ref{sec:results}, future work could examine wider classes of
statistics-preserving transformations; for example, anything of the
form
\begin{equation}
\delta(\vk) \to T(\vk) \delta(\vk) \label{eq:transformation-T}
\end{equation}
with $|T(\vk)|^2=1$ is suitable. For the overdensity field to remain
real, one additionally requires $T(\vk)=T(-\vk)^{\star}$ but otherwise there are no
restrictions. In particular there is no requirement for $T$ to be
isotropic or homogeneous. 

Condition \eqref{eq:transformation-T} ensures that the power spectrum
is unchanged; our method of cross-correlation will then allow the
study of structure growth in the presence of the correct cosmological
background. The form of $T(\vk)$ is dictated by the specific aspect of
structure growth under study.
 
Section \ref{sec:results}'s A-B simulations correspond to the simplest
case of $T(\vk)=-1$; as another example, a translation
$\vx \to \vx + \Delta \vx$ corresponds to the case
$T(\vk) = e^{i\vk\cdot\Delta\vx}$. Let us briefly consider one further
illustrative extension, given by
\begin{equation}
T(\vk) = 1-2\Theta\left(k_0-|\vk|\right)\textrm{,}
\end{equation}
where $\Theta$ is the Heaviside step function. The resulting
transformation flips the sign of $\delta$ for wavenumbers below a
critical $k_0$. We can refer to the simulation resulting from the new
initial conditions as `spliced' (abbreviated to S) since the initial
conditions are identical to $A$ for $k>k_0$ and to $B$ for $k<k_0$. In
terms of perturbation theory, this splicing operation is more complex
than a $k$-independent transformation because it breaks loop terms
into an infrared and ultraviolet part with different signs. Being able to 
segregate parts of loop integrals fully within a numerical simulation 
in principle allows a very detailed comparison with perturbation theory. 
Here we will consider only the qualitative results.

Because the S and A simulations are anticorrelated on large scales,
the low-$k$ modes destroy the high-$k$ correlations over time
(Fig.~\ref{fig:spliced}, top panel) just as with the A-B cross
correlation (Fig.~\ref{fig:cross-ps}, lower panel). On the other hand,
cross-correlating S with B reveals that the low-$k$ modes remain
positively correlated at all times, showing that the anticorrelation on
small scales does not affect the larger scales. Furthermore, as
non-linear power grows in the late-time universe, the ratio $P_{SB}/P$
ultimately becomes positive at large $k$: structure
growth is coherent between the S and B universes because it is
regulated by the largest scale modes.  A full understanding of the
coupling of large- and small-scale modes is necessary for
distinguishing the bispectrum due to non-linear evolution
\cite{2015MNRAS.448L..11W} from any primordial contribution. Using our technique this
behavior is exposed to quantitative study without ever changing the
power spectrum away from $\Lambda$CDM, and with just three simulations
rather than expensive averages over large numbers
\cite{1999MeiksinWhiteCorrelatedPk}.

One could expand to an even broader class of
transformations where two independent, uncorrelated initial
realizations ($\delta_{\mathrm{I}}$ and $\delta_{\mathrm{II}}$, say)
are available:
\begin{equation}
\delta(\vk) \to T(\vk) \delta_{\mathrm{I}}(\vk) + S(\vk) \delta_{\mathrm{II}}(\vk)\label{eq:most-general-transformation}
\end{equation}
where $|T(\vk)|^2 + |S(\vk)|^2=1$. This includes another interesting
special case where the $k<k_0$ modes are kept fixed while $k>k_0$
modes are randomized (rather than anticorrelated). This specialization
has been studied elsewhere \cite{2012AragonCalvoMIP} to average away
stochastic fluctuations in halo spin alignments
\cite{2014AragonCalvoSpinAlignmentsMIP} and local bias measurements
\cite{2014NeyrinckMIPbias}, using large numbers of runs with
independent $\delta_{\mathrm{II}}$ fields. Our approach of pulling out
information from a single additional simulation with transformed
initial conditions could also be applied to this specialization, for
example as another way to isolate the contribution of specific $k$
modes to the perturbation theory loop terms.

\section{Conclusions}\label{sec:conclusions}

We have introduced the technique of ``paired'' simulations. We run two
simulations (``A'' and ``B'') that are identical except for having
inverted initial linear overdensities ($\delta_A = - \delta_B$). Since
by definition the linear field is symmetric about zero, the two
simulations have identical statistical properties. We illustrated how
this can be used to better understand the evolution of voids; extract
information on the physical basis of perturbation theories; and eliminate 
a class of finite-volume effects from power spectrum estimates with 
greater efficiency than existing techniques. Extensions to a broader
class of transformations of the initial density field could further
enhance the power of this technique.

\acknowledgments AP acknowledges financial support from the Royal
Society. NR and HVP were supported by the European Research Council
under the European Community's Seventh Framework Programme (FP7/2007-
2013) / ERC grant agreement no 306478-CosmicDawn.  AS thanks the
Department of Physics and Astronomy at University College London for
hospitality during completion of this work.  HVP thanks the Galileo
Galilei Institute for Theoretical Physics for hospitality and the INFN
for partial support during the completion of this work. This work used
the DiRAC Complexity system, operated by the University of Leicester
IT Services, which forms part of the STFC DiRAC HPC Facility
(\url{www.dirac.ac.uk}). This equipment is funded by BIS National
E-Infrastructure capital grant ST/K000373/1 and STFC DiRAC Operations
grant ST/K0003259/1. DiRAC is part of the National E-Infrastructure.

\bibliographystyle{apsrev}
\bibliography{refs.bib}

\appendix

\section{AB cross-correlations in the Zel'dovich approximation}\label{sec:zeldovich-cross-corr}

In this Appendix, we derive the result
quoted in the text for the cross-power spectrum of the A and B
simulations in the Zel'dovich approximation. Our approach closely
follows that of previous works \cite{1996TaylorHamiltonLPT,Matsubara08LPT}, but
extends to the cross-power spectrum with an alternative resummation that
we will describe in due course.

The Zel'dovich approximation is the linear-order solution to
Lagrangian perturbation theory. The central quantity is the
displacement field $\vPsi(\vq)$ which describes the movement of
particles from their initial positions $\vq$ to their final position
$\vx=\vq+\vPsi(\vq)$. All information about the system is then
expressed in terms of $\vPsi(\vq)$. For example, the local density is
\begin{equation}
\rho(\vx) = \bar{\rho} \left| \frac{\dd^3 \vec{q}}{\dd^3 \vec{x}} \right|\textrm{,}\label{eq:density-from-displacement}
\end{equation}
where $|\dd^3 \vec{q} / \dd^3 \vec{x}|$ denotes the Jacobian
determinant of the transformation between Lagrangian and Eulerian
coordinates $\vec{q}$ and $\vec{x}$, and $\bar{\rho}$ is the
volume-averaged density. This allows the fractional overdensity in
Fourier space, $\delta(\vk)$, to be written
\begin{align}
  \delta(\vk) & \equiv \int \dd^3 \vec{x} \,
  \left(\frac{\rho(\vec{x})}{\bar{\rho}} - 1 \right)\, e^{-i\vk\cdot
                \vx} \nonumber \\
& = \int \dd^3 \vq\, e^{-i\vk\cdot\vq} \left[e^{-i\vk\cdot\vPsi(\vq)} -1 \right]\textrm{,}\label{eq:k-overdensity-from-displacement}
\end{align}
where expression \eqref{eq:density-from-displacement} has been used to
transform the integration variable to $\vec{q}$ for the first term,
whereas the second term has been rewritten with a relabeling of
the integration coordinate from $\vec{x}$ to $\vec{q}$. Mass
conservation demands that $\langle \delta(\vk) \rangle = 0$ which,
combined with Eq.~\eqref{eq:k-overdensity-from-displacement},
implies
\begin{equation}
\int \dd^3 \vq e^{-i \vk \cdot \vq} \langle e^{-i \vk \cdot
  \vPsi(\vq)} \rangle = \int \dd^3 \vq e^{-i \vk \cdot \vq}\textrm{,}\label{eq:mean-delta-0}
\end{equation}
a result that we will use momentarily.

The cross-power spectrum $P_{XY}(k)$ between fields $X$ and $Y$ is defined by
\begin{equation}
\langle \delta_X(\vec{k}) \delta_Y(\vec{k'}) \rangle = (2 \pi)^3 \delta^D(\vec{k} + \vec{k}') P_{XY}(k) \textrm{,}
\end{equation}
where $\delta^D$ is the Dirac delta-function. Substituting two
copies of expression \eqref{eq:k-overdensity-from-displacement} into
this definition gives the following expression for the
cross-power in terms of the displacement fields:
\begin{equation}
  P_{XY}(k) = \int \dd^3 \vr\, e^{-i\vk\cdot\vr} \left[\left<
   e^{-i\vk\cdot \Delta \vPsi} \right> -1 \right],\label{eq:power-spectrum-from-displacement}
\end{equation}
where we have used Eq.~\eqref{eq:mean-delta-0} and defined
$\Delta \vPsi \equiv \vPsi_X(\vq) - \vPsi_Y(\vq')$ with $\vr\equiv\vq-\vq'$.

The treatment to this point has been exact (up to shell crossing).  We
now introduce the perturbative element by employing the Zel'dovich
approximation, in which the displacement is related to the
linear-theory density field $\delta_L(\vk)$ by
\begin{equation}
  \vPsi_X^{\mathrm{zel}} (\vk) = i \alpha_X \frac{\vk}{k^2} \delta_{L}(\vec{k})\label{eq:zeldovich}
\end{equation}
for a constant $\alpha_X$ depending on which field
$X$ we consider. For the A field, $\alpha_A=1$; for the B field,
$\alpha_B=-1$.

We additionally want to be able to calculate the cross-correlation
$P_{AL}$ between the true and linearly-evolved fields. Since the
Zel'dovich approximation and linear theory have to agree in the limit
of small density variations, we can represent the linear theory by
multiplying the displacements by some small number $\alpha_L \ll 1$, but
then dividing the output density field
\eqref{eq:k-overdensity-from-displacement} by the same small number
$\alpha_L$. One can think of this procedure as rescaling the input
linear field by a growth factor appropriate for some very early
time, then undoing the scaling in the final expression. As a
verification, we can re-calculate the linear field
from Eq.~\eqref{eq:k-overdensity-from-displacement}, substituting
\eqref{eq:zeldovich} and then Taylor expanding in $\alpha_L$:
\begin{align}
\delta^{\mathrm{zel}}_L(\vk) & \simeq \frac{1}{\alpha_L} \int \dd^3 \vec{q}\, e^{-i \vk
  \cdot \vq} \int \frac{\dd^3 \vk'}{(2 \pi)^3} \alpha_L
                               \frac{\vk \cdot \vk'}{\vk'^2}
                               \delta_L(\vk') e^{i \vk' \cdot \vq} \nonumber \\
& = \delta_L(\vk)\textrm{,}
\end{align}
confirming the recovery of the linear density field.

We now return to Eq.~\eqref{eq:power-spectrum-from-displacement}
and insert expression~\eqref{eq:zeldovich} for the displacement
fields.  In the Zel'dovich picture, $\vPsi_X$ is always Gaussian, so
we can apply the cumulant expansion
$\left<e^{-G}\right> = \exp\left[-\frac{1}{2} \left <G^2\right>
\right]$.
We divide the final expression by $|\alpha_X \alpha_Y|$; for
$\alpha_A=1$ and $\alpha_B=-1$ this has no effect, whereas for
$\alpha_L \ll 1$ this captures the shift to linear theory described
above.  Put together, we obtain the following expression for auto- and
cross-spectra:
\begin{equation}
P^{\mathrm{zel}}_{XY}(k) = \frac{1}{\left|\alpha_X \alpha_Y
  \right|}\int \dd^3r e^{-i\vk\cdot\vr}
\left[ e^{-(\alpha_X-\alpha_Y)^2 I(\vk,0)/2 + \alpha_X \alpha_Y J(\vk,\vr)} - 1\right] \label{eq:lpt1}
\end{equation}
where $I(\vk,\vr)$ captures the effect of the displacement field for
two points at distance $\vr$ on a Fourier mode with wavevector $\vk$
and is given by
\begin{equation}
  I(\vk,\vr) = \int \frac{\dd^3 \vk'}{(2\pi)^3} \frac{(\vk \cdot \vk')^2}{k'^4}
  \cos(\vk' \cdot \vr)\,P_L(k')\textrm{,}
\end{equation}
and $J(\vk,\vr) = I(\vk,\vr)-I(\vk,0)$. The exponent in the integral above has two pieces, one that depends on $\vr$ (proportional to $J(\vk,\vr)$) and the one that does not (proportional to $I(\vk,0)$). The piece that does not depend on $\vr$ can be pulled out of the integral to generate a $k$-dependent exponential suppression. The exponential for the other piece is expanded to first order and integrated to generate a linear power spectrum term and a harmless $k=0$ correction
\begin{align}
P^{\mathrm{zel}}_{XY}(k) &= e^{-(\alpha_X-\alpha_Y)^2(k/\kNL)^2} \left[P_L(k)-I(\vk,0)\delta^D(\vk)\right]\textrm{,}\label{eq:PXY_zel}
\end{align}
where  $\kNL$ is the wavenumber corresponding to a nonlinearity
scale, defined by 
\begin{align}
\kNL^{-2} &= \frac{1}{2} I(k,0)\,k^{-2} = \frac{1}{12\pi^2} \int_0^{\infty} P_L(k') \dd k'\textrm{.}
\end{align}
It follows that
\begin{eqnarray}
  P^{\mathrm{zel}}_{AL}(k) &=& e^{-(k/\kNL)^2} P_L(k) \, \textrm{, and } \\
  P^{\mathrm{zel}}_{AB}(k) &=& e^{-4(k/\kNL)^2} P_L(k) \, , \label{eq:zel-AB-our-resummation}
\end{eqnarray}
which are the results quoted in the main text. The decorrelation scale between the A and B fields is half that of the decorrelation scale between A and L fields, a result confirmed in our measurements from the simulations.

This derivation looks deceptively close to the resummed LPT approach
of Ref.~\cite{Matsubara08LPT}, which we denote by ``M''. However,
there is an important conceptual difference in the detail. In Ref.~\cite{Matsubara08LPT}, the equivalent of our Eq.~\eqref{eq:lpt1} is given by
\begin{equation}
P^{\mathrm{zel,M}}_{XY}(k) = \frac{1}{\left|\alpha_X \alpha_Y
  \right|}\int \dd^3r e^{-i\vk\cdot\vr}
\left[ e^{-(\alpha_X^2+\alpha_Y^2) I(\vk,0)/2 + \alpha_X \alpha_Y I(\vk,\vr)} - 1\right] \label{eq:lpt1m}.
\end{equation}
Expressions \eqref{eq:lpt1} and \eqref{eq:lpt1m} are mathematically equivalent: we have just moved a constant from the $\vr$-independent piece back into the $\vr$-dependent piece (which we later expand). At infinite order in the expansion this does not matter, but following the same reasoning as above we obtain to the first order in expanded integral,
\begin{eqnarray}
  P^{\mathrm{zel,M}}_{AL}(k) &=& e^{-(k/\kNL)^2} P_L(k), \\
  P^{\mathrm{zel,M}}_{AB}(k) &=& e^{-2(k/\kNL)^2} P_L(k). 
\end{eqnarray}
The cross-correlation between the initial and final fields is the
same, but the AB decorrelation scale differs by a factor of two. The
difference can be attributed to re-summation of a different set of
operators, demonstrating the fragility of this operation.  Equation
\eqref{eq:lpt1} is constructed such that the perturbative part
vanishes for effects arising at small separations, $r \to 0$.  This is
why it agrees with the intuitive description given in Section
\ref{sec:pert-theory} that the AB particle displacement is doubled
relative to the AL displacement --- this argument implicitly assumes
that the displacements are coherent, which only becomes exactly true
in the $r \to 0$ (``eikonal'') limit.  Conversely,
Eq.~\eqref{eq:lpt1m} is ideally suited to expanding the auto-power
($\alpha_X = \alpha_Y = 1$) but has no special properties as
$r \to 0$. A more systematic understanding of these differences is
beyond the scope of this work.

\vspace{2cm}

\end{document}